\documentclass[fleqn,usenatbib,useAMS]{mnras}
 \usepackage{graphicx}
 \usepackage{amsmath}
 \usepackage{amssymb}
 \usepackage[T1]{fontenc}
 \usepackage{ae,aecompl}
 \usepackage{txfonts}
 \usepackage{enumerate}

 \title[Searching for the lost Unicorn in Vela]
       {Searching for the lost Unicorn: a prominent feature in the radial
        velocity distribution of stars in Vela from \textit{\textbf{Gaia}} DR2 data}

 \author[R. de la Fuente Marcos and C. de la Fuente Marcos]
        {R.~de~la~Fuente Marcos$^{1}$\thanks{E-mail: rauldelafuentemarcos@ucm.es}
         and
         C. de la Fuente Marcos$^2$ \\
         $^1$AEGORA Research Group,
             Facultad de Ciencias Matem\'aticas,
             Universidad Complutense de Madrid,
             Ciudad Universitaria, E-28040 Madrid, Spain \\
         $^2$ Universidad Complutense de Madrid,
              Ciudad Universitaria, E-28040 Madrid, Spain}
 \date{Accepted 2018 August 29.
       Received 2018 August 29;
       in original form 2018 July 8}
 \pubyear{2018}
 
\begin{document}
  \label{firstpage}
  \pagerange{\pageref{firstpage}--\pageref{lastpage}}
  \maketitle

  \begin{abstract}
     Stellar streams are ubiquitous in the Galactic halo and they can be used 
     to improve our understanding of the formation and evolution of the Milky
     Way as a whole. The so-called Monoceros Ring might have been the result 
     of satellite accretion. Guglielmo et al. have used $N$-body simulations 
     to search for the progenitor of this structure. Their analysis shows 
     that, if the Ring has a dwarf galaxy progenitor, it might be found in 
     the background of one out of eight specific areas in the sky. Here, we 
     use {\it Gaia} DR2 data to perform a systematic exploration aimed at 
     confirming or rejecting this remarkable prediction. Focusing on the 
     values of the radial velocity to uncover possible multimodal spreads, 
     we identify a bimodal Gaussian distribution towards Galactic coordinates 
     ($l$, $b$) = (271{\degr}, +2{\degr}) in Vela, which is one of the 
     locations of the progenitor proposed by Guglielmo et al. This prominent 
     feature with central values 60$\pm$7~km~s$^{-1}$ and 
     97$\pm$10~km~s$^{-1}$, may signal the presence of the long sought 
     progenitor of the Monoceros Ring, but the data might also be compatible 
     with the existence of an unrelated, previously unknown, kinematically 
     coherent structure.   
  \end{abstract}


  \section{Introduction}
     Beyond the nominal edge of the Milky Way disc, 15~kpc from the Galactic centre, lies a complex network of coherent stellar structures
     whose origins are not yet fully understood (see e.g. \citealt{2011PhDT.......171S,2011ApJ...738...79X,Pila15,2016ASSL..420..113S}). 
     Some may be the result of the Milky Way cannibalizing nearby dwarf galaxies \citep{1994Natur.370..194I}, others could be just stellar 
     lumps induced by the combined action of the gravitational potential of the Galaxy and those of passing and/or falling neighbours 
     \citep{2015ApJ...801..105X,2018MNRAS.478.3809S}. The study of these structures can help in understanding how the Milky Way came into
     existence and how it has evolved progressively to become what we observe now. Among all these structures, the true nature of the 
     so-called Monoceros Ring remains elusive. Originally identified by \citet{2002ApJ...569..245N}, the structure is considered by some as 
     a {\it bona fide} stellar stream (e.g. \citealt{2003MNRAS.340L..21I,2003ApJ...588..824Y,2007MNRAS.376..939C,2008ApJ...689L.117G,
     2008ApJ...684..287I,2011MNRAS.414L...1M,2011ApJ...730L...6S,2018MNRAS.473.1218L}), while others put its origin in the (flared thick) 
     disc of the Milky Way (e.g. \citealt{2005ApJ...630L.153C,2006A&A...451..515M,2014ApJ...794...90K,2014A&A...567A.106L,2018ApJ...854...47S,
     2018MNRAS.478.3367W}). Alternative scenarios put its provenance in an outer spiral arm such as the one described by 
     \citet{2011ApJ...734L..24D} or in undulations of the disc \citep{2017ApJ...844...74L} like those discussed by \citet{2015ApJ...801..105X}.
    
     \citet{2018MNRAS.474.4584G} have recently explored the kinematics, proper motions, and the nature of the putative progenitor of the 
     Monoceros Ring. Although they could not confirm that the Ring has its origin in an accretion episode, their analysis argues that, if 
     the Ring has a dwarf galaxy progenitor, it might be found in the background of one out of eight well-defined areas in the sky. Here, we 
     use {\it Gaia} DR2 data to perform a systematic exploration of these areas aimed at confirming or rejecting their prediction, focusing 
     on the values of the radial velocity to uncover possible multimodal distributions. This Letter is organized as follows. Section~2 
     presents the input data and tools used in our analysis. The eight radial velocity distributions are explored in Section~3. Section~4 
     presents the analysis of a bimodal Gaussian distribution found towards Galactic coordinates ($l$, $b$) = (271{\degr}, +2{\degr}) in 
     Vela. In Section~5, we study the statistical significance of our findings. Results are discussed in Section~6 and conclusions are 
     summarized in Section~7.

  \section{Input data and tools}
     \citet{2018MNRAS.474.4584G} have proposed that, if the Monoceros Ring has a dwarf galaxy progenitor, it might be found towards one out
     of eight patches of sky. In their fig. 2, three patches are linked to their Model 2, two to their Model 3, and three more to their 
     Model 4. Model 2 predicts a progenitor located in the range of Galactocentric distances, $d$, 27$\pm$12~kpc, Model 3 predictions are 
     for the range 40$\pm$12~kpc, and those from Model 4 span the range 31$\pm$13~kpc. Model 2 indicates three possible locations centred at 
     Galactic coordinates ($l$, $b$) of (14{\degr}, $-$1{\degr}) in the constellation of Sagittarius (henceforth Model 2 A), (352{\degr}, 
     $-$2{\degr}) in Scorpius (Model 2 B), and (232{\degr}, +2{\degr}) in Puppis (Model 2 C); Model 3 gives (11{\degr}, $-$2{\degr}) in
     Sagittarius (Model 3 A) and (354{\degr}, $-$1{\degr}) in Scorpius (Model 3 B); Model 4 suggests (17{\degr}, $-$2{\degr}) in Scutum 
     (Model 4 A), (249{\degr}, $-$1{\degr}) in Puppis (Model 4 B), and (271{\degr}, +2{\degr}) in Vela (Model 4 C). Our objective is to 
     analyse data in a three-dimensional sub-space (eight of them) constrained by the values of $l$, $b$ and $d$ in \citet{2018MNRAS.474.4584G}.   

     {\it Gaia} DR2 \citep{2016A&A...595A...1G,2018A&A...616A...1G} provides extensive astrometric and photometric data ---namely, 
     coordinates, parallax, radial velocity, proper motions, blue, red and green magnitudes, and their respective standard errors--- that 
     can be used to perform the exploration required to confirm or reject the prediction made by \citet{2018MNRAS.474.4584G}. Out of 
     87\,733\,672 sources with strictly positive values of the parallax, we found 4\,831\,766 sources with values of the radial velocity,
     $V_r$. This is our primary sample and we have been able to extract 475, 878, 102, 252, 247, 344, 199, and 150 sources, respectively, 
     for the eight patches of sky singled out by \citet{2018MNRAS.474.4584G}. The extraction process has been restricted to a square region 
     of side 10{\degr}, centred at the proposed locations; i.e. for Model 2 A, the ranges in $l$ and $b$ are, respectively, (9{\degr}, 
     19{\degr}) and ($-$6{\degr}, +4{\degr}) with distances within 27$\pm$12~kpc. We believe that the size of our sampling window is large 
     enough to obtain robust results (see fig. 2 in \citealt{2018MNRAS.474.4584G}).
   
     When needed and following the approach outlined by \citet{1987AJ.....93..864J}, we have used the input data to compute Galactic space 
     velocities and their uncertainties in a right-handed coordinate system for $U$, $V$, and $W$; axes are positive in the directions of 
     the Galactic centre, Galactic rotation, and the North Galactic Pole (NGP). The Galactocentric standard of rest is a right-handed 
     coordinate system centred at the Galactic centre. Axes are positive in the directions of the Sun, Galactic rotation, and the NGP. The 
     necessary values of the Solar motion were taken from \citet{2010MNRAS.403.1829S}. Averages, standard deviations, medians, interquartile 
     ranges (IQRs) and other statistical parameters have been computed in the usual way (see e.g. \citealt{2012psa..book.....W})

  \section{Velocity distributions}
\begin{figure}
        \centering
         \includegraphics[width=\linewidth]{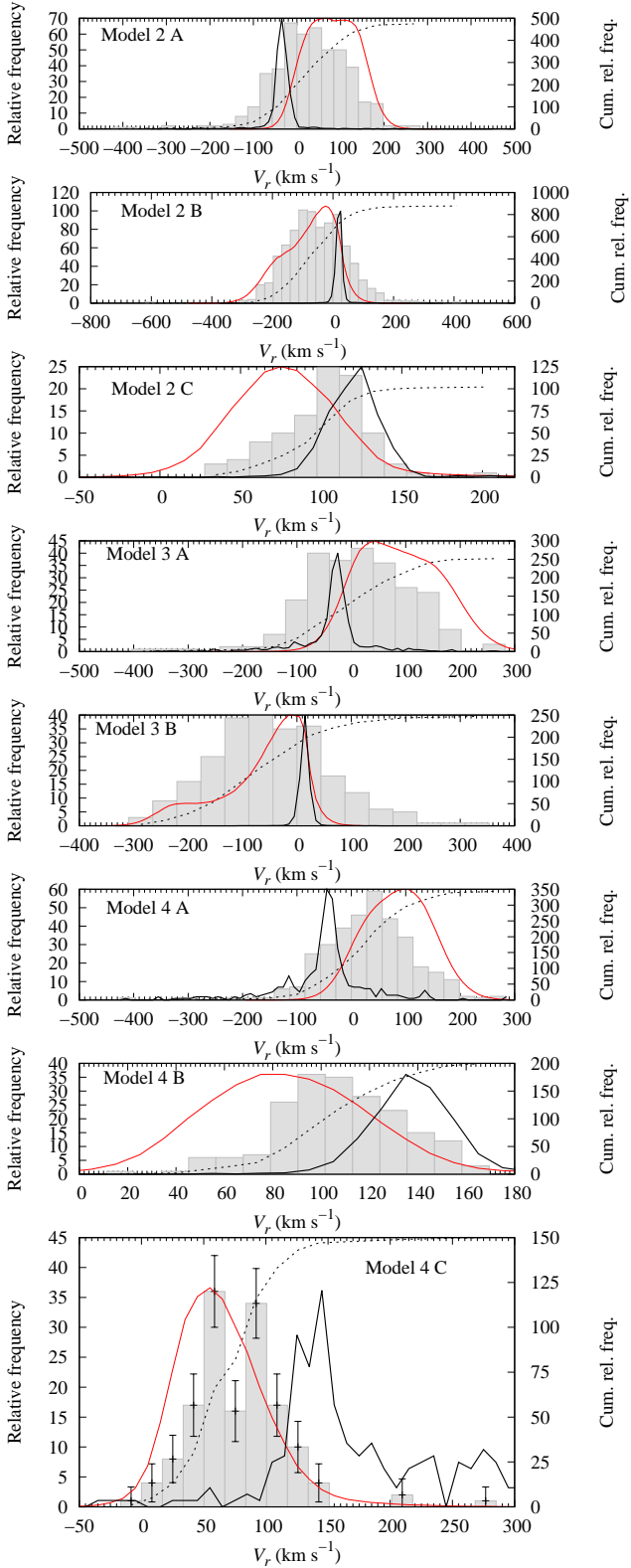}
         \caption{Radial velocity distributions of the {\it Gaia} DR2 samples associated with the eight possible progenitor locations of the 
                  Monoceros Ring discussed in \citet{2018MNRAS.474.4584G}. Solid line distributions from \textsc{galaxia} 
                  \citep{2011ApJ...730....3S} with warp and flare. See the text for details. 
                 }
         \label{vrdist}
      \end{figure}

     Fig.~\ref{vrdist} shows the radial velocity distributions of the eight {\it Gaia} DR2 samples associated with the possible locations of
     the progenitor of the Monoceros Ring discussed by \citet{2018MNRAS.474.4584G}. The histograms displayed here have a statistically 
     meaningful bin width computed using the Freedman-Diaconis rule \citep{FD81}, i.e. $2\ {\rm IQR}\ n^{-1/3}$, where $n$ is the number of 
     sources; cumulative frequencies are plotted as dashed curves. The values of the IQRs (median values in parentheses) in km~s$^{-1}$ are, 
     respectively (top to bottom in Fig.~\ref{vrdist}), 110.27 (+31), 134.93 ($-$56.98), 32.51 (+105.22), 127.33 (+15.82), 138.47 ($-$58.01), 
     99.33 (+31.83), 32.99 (+106.14), and 44.56 (+75.78), and the bin sizes of the associated histograms are, respectively (also in 
     km~s$^{-1}$), 28.26, 28.18, 13.92, 40.31, 44.14, 28.35, 11.30, and 16.77. Given the fact that the values of the IQRs for the different 
     regions are quite different, our analysis is sensitive to finding kinematically cold substructures only in a few of them. 

     As the Galactic latitude of the various samples corresponds to the disc, the presence of a single peak in the radial velocity 
     distribution can be interpreted as the signature of a disc-like population; if additional peaks are found, they may correspond to 
     coherent structures different from the disc. As a reference, the solid lines in Fig.~\ref{vrdist} show the distributions of synthetic 
     samples of field stars generated from the \textsc{galaxia}\footnote{\url{http://www.physics.usyd.edu.au/~sanjib/code/}} model 
     \citep{2011ApJ...730....3S} within one square degree around the region's centre (thin red curve, entire synthetic sample, thick black 
     curve, synthetic sample with Galactocentric distance $>$ 15~kpc, normalized to the peak of the corresponding histogram). In general and 
     with the exception of the sample of location C of Model 4 in \citet{2018MNRAS.474.4584G} ---see Fig.~\ref{vrdist}, bottom panel--- it 
     is not possible to identify any statistically significant bimodal behaviour. The error bars in the bottom panel of Fig.~\ref{vrdist} 
     have been calculated using Poisson statistics ($\sigma=\sqrt{n}$) and applying the approximation given by \citet{1986ApJ...303..336G} 
     when $n<21$, $\sigma \sim 1 + \sqrt{0.75 + n}$.  
%
%
%
%

  \section{A bimodal distribution of radial velocities in Vela}
\begin{figure}
        \centering
         \includegraphics[width=\linewidth]{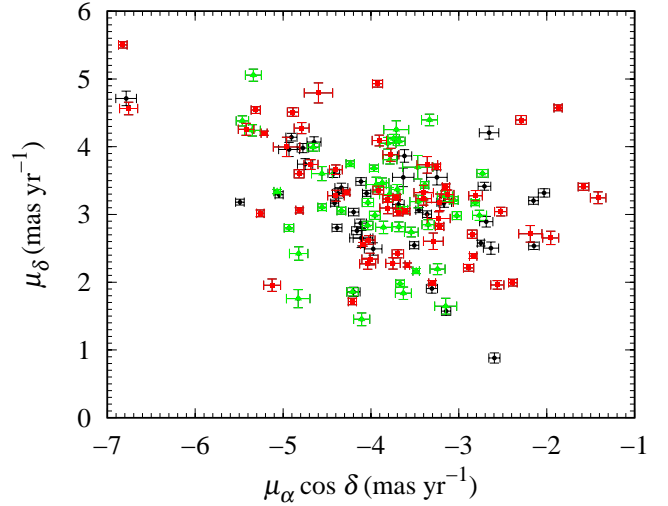}
         \caption{Proper motion components with 1$\sigma$ error bars for the {\it Gaia} DR2 sample of location C of Model 4 in 
                  \citet{2018MNRAS.474.4584G}: green triangles, stars assumed to be part of the disc (45~km~s$^{-1}$$<V_r<$ 75~km~s$^{-1}$), 
                  red squares, stars linked to the high-velocity peak in the bottom panel of Fig.~\ref{vrdist} (80~km~s$^{-1}$$<V_r<$ 
                  120~km~s$^{-1}$), others plotted as black circles.
                 }
         \label{pm}
      \end{figure}

\begin{figure}
        \centering
         \includegraphics[width=\linewidth]{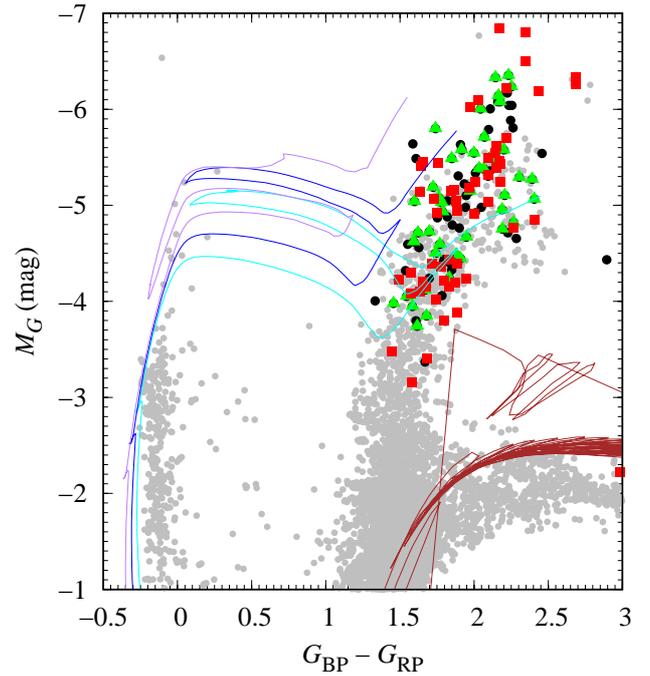}
         \caption{Colour-magnitude diagram for the same sample in Fig.~\ref{pm}, similar to fig. 5 in \citet{2018A&A...616A..10G}. The 
                  entire {\it Gaia} DR2 sample of location C of Model 4 is plotted in light grey. Three
                  isochrones of an age of 50~Myr and solar (cyan), solar/5 (blue), and solar/50 (purple) metallicities and one of an age of 
                  1.5~Gyr (brown) and solar metallicity from \citet{2017ApJ...835...77M} are plotted as a reference.
                 }
         \label{cmd}
      \end{figure}

     The bottom panel in Fig.~\ref{vrdist} seems to follow a bimodal Gaussian distribution produced by populations with central velocities
     60$\pm$7~km~s$^{-1}$ and 97$\pm$10~km~s$^{-1}$ and consistent dispersions of about 10~km~s$^{-1}$. Assuming that the observed 
     distribution is mainly the sum of two normal distributions, the bimodality index (difference of means divided by dispersion) is 3.7, 
     and the ratio of the two populations is nearly 1:1. We interpret the population with the lowest central velocity as an extension of 
     the thin disc at average Galactocentric distances close to 30~kpc. This interpretation appears to be compatible with predictions from
     \textsc{galaxia}, but the higher $V_r$ peak is neither compatible with the thin disc (thin curve) nor with the halo component (thick 
     curve), which should have values close to 150~km~s$^{-1}$. Fig.~\ref{pm} shows that both populations (low velocity in green, high 
     velocity in red) have similar distributions in terms of proper motions ($\mu_{\alpha} \cos\delta$, $\mu_{\delta}$), although the 
     high-velocity group exhibits a larger dispersion in the values of their proper motions in right ascension ---averages, standard 
     deviations, medians, and IQRs are (disc members) $-$3.9$\pm$0.7~mas~yr$^{-1}$, $-$3.78~mas~yr$^{-1}$, 0.77~mas~yr$^{-1}$ (right 
     ascension), 3.2$\pm$0.8~mas~yr$^{-1}$, 3.19~mas~yr$^{-1}$, 0.88~mas~yr$^{-1}$ (declination), and (higher $V_r$ peak) 
     $-$3.8$\pm$1.1~mas~yr$^{-1}$, $-$3.73~mas~yr$^{-1}$, 1.24~mas~yr$^{-1}$ (right ascension), 3.3$\pm$0.9~mas~yr$^{-1}$, 
     3.24~mas~yr$^{-1}$, 1.11~mas~yr$^{-1}$ (declination). As the values of their distances span the same range ---averages, standard 
     deviations, medians, and IQRs for the parallaxes are (disc) 0.039$\pm$0.009~mas, 0.039~mas, 0.018~mas, and (higher $V_r$ peak) 
     0.040$\pm$0.011~mas, 0.040~mas, 0.019~mas--- their tangential velocities are also similar. The colour-magnitude diagram in 
     Fig.~\ref{cmd} displays the intrinsic values (i.e. corrected for extinction and reddening using the data provided by {\it Gaia} DR2) of 
     these two populations, which follow similar distributions made of probable young stars, perhaps classical Cepheids, although the 
     high-velocity population seems to be slightly older and/or have different metallicity. Fig.~\ref{cmd} is similar to figs~19b and 20 in 
     \citet{2018A&A...616A...8A} in which the red giant branch (RGB) corresponds to an age of 1--2~Gyr. As a reference, three 50 Myr and one 
     1.5 Gyr PARSEC v1.2S + COLIBRI PR16\footnote{\url{http://stev.oapd.inaf.it/cgi-bin/cmd}} isochrones are also plotted. Most sources in 
     the higher $V_r$ peak of the bottom panel of Fig.~\ref{vrdist} are too bright to be RGB stars.

%
%
%
%
%
%
%
%

  \section{Statistical significance}
\begin{figure}
        \centering
         \includegraphics[width=\linewidth]{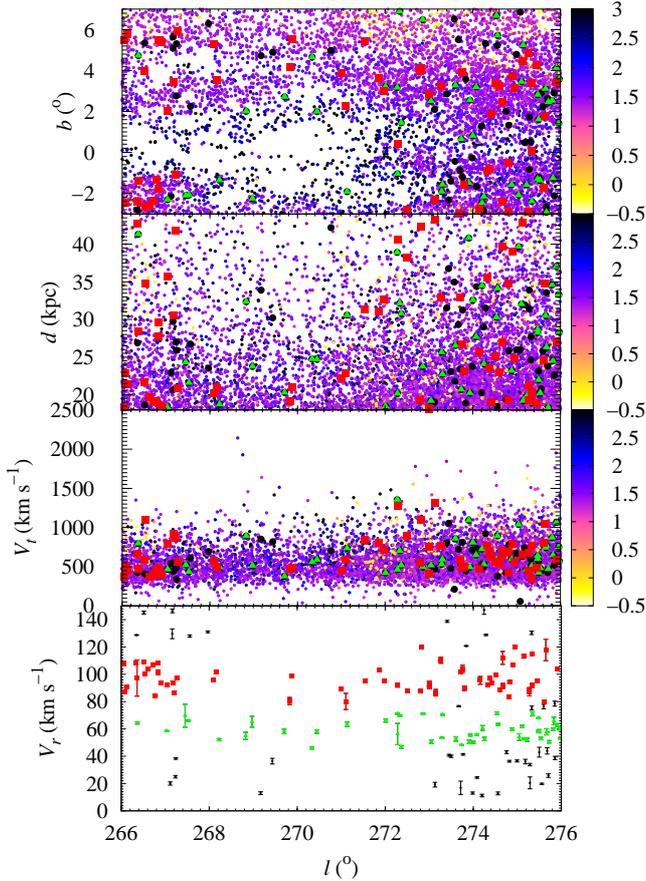}
         \caption{Distribution in Galactic coordinates (top panel), distance (second to top panel), and tangential velocity (second to 
                  bottom panel) for the entire {\it Gaia} DR2 sample of location C of Model 4 in \citet{2018MNRAS.474.4584G}. The colour map 
                  shows the value of the colour index in Fig.~\ref{cmd}, computed as shown in \citet{2018A&A...616A..10G}. Hot stars appear 
                  lighter in colour and cold stars darker. Points in green and red as in Figs~\ref{pm} and \ref{cmd}. The radial velocity
                  distribution with error bars is shown in the bottom panel.
                 }
         \label{full}
      \end{figure}

     Fig.~\ref{vrdist}, bottom panel, shows that the sample from Model 4 C is bimodal with central values 60$\pm$7~km~s$^{-1}$ and 
     97$\pm$10~km~s$^{-1}$. The separation between the peaks is 34~km~s$^{-1}$, which is of order of IQR and well above the usual values of 
     the uncertainties in the values of the radial velocities (well below 5~km~s$^{-1}$ in most cases, see Fig.~\ref{full}, bottom panel). 
     The difference between the bin centred at 58.7~km~s$^{-1}$ and the one at 75.5~km~s$^{-1}$ is nearly 4$\sigma$; in addition, the excess 
     of the bin centred at 92.2~km~s$^{-1}$ with respect to the one at 75.5~km~s$^{-1}$ is about 3.5$\sigma$ ---in both cases using the 
     $\sigma$-value at the dip. If we repeat the same analysis for Model 2 B (Fig.~\ref{vrdist}, second to top panel), the single other that 
     might exhibit signs of bimodality, excesses of just 1.1$\sigma$ are found.
%
%
%
%

  \section{Discussion}
     In general, the data from {\it Gaia} DR2 for sources beyond 15~kpc are affected by large uncertainties, particularly large in the case
     of the parallax values (i.e. distances). Our main findings are based on a parameter, the value of the radial velocity, which is 
     definitely the least uncertain of the data set, if measured. The samples used in this investigation have very good values of the radial
     velocity and also the proper motions, but the values of the distances derived from the parallaxes are rather uncertain. The values of
     the tangential velocities and absolute magnitudes inherit these large errors. Therefore, our conclusions must be based on the value of 
     the radial velocity and those of the components of the proper motion. These are sufficiently reliable.

     It may be argued that both peaks in the distribution of radial velocity in Fig.~\ref{vrdist}, bottom panel, could be the result of the
     presence of small, distant, and decaying clusters such as E 3 \citep{2015A&A...581A..13D}. In such cases, we should observe very small 
     concentrations of relevant sources. Fig.~\ref{full}, top panel, shows the distribution of low- and high-velocity sources in Galactic 
     coordinates and no such obvious concentrations are observed. The same can be said about the values of the distance in Fig.~\ref{full}, 
     second to top panel. In Fig.~\ref{full}, both the sample with radial velocities and the most general one are plotted using the 
     intrinsic colour index in Fig.~\ref{cmd} as third coordinate. Fig.~\ref{full}, top panel, clearly indicates that the effects of 
     extinction are very important in this region with the empty spaces corresponding to putative missing sources not present in {\it Gaia} 
     DR2 because the foreground population and associated gas clouds effectively block the light from these sources. The fact that the 
     sources surrounding the empty spaces tend to have the largest reddening values, clearly confirms that reddening is a major concern 
     here. It is however unlikely that the dominant role that interstellar extinction indeed has in this region may have a major impact on 
     our conclusions, which are based on the distribution of radial velocities for the most part. Fig.~\ref{full}, second to bottom panel, 
     shows the values of the tangential velocity, $V_t$, which exhibit a considerable spread due to the large uncertainties in the values of 
     the distances.

\begin{figure}
        \centering
         \includegraphics[width=\linewidth]{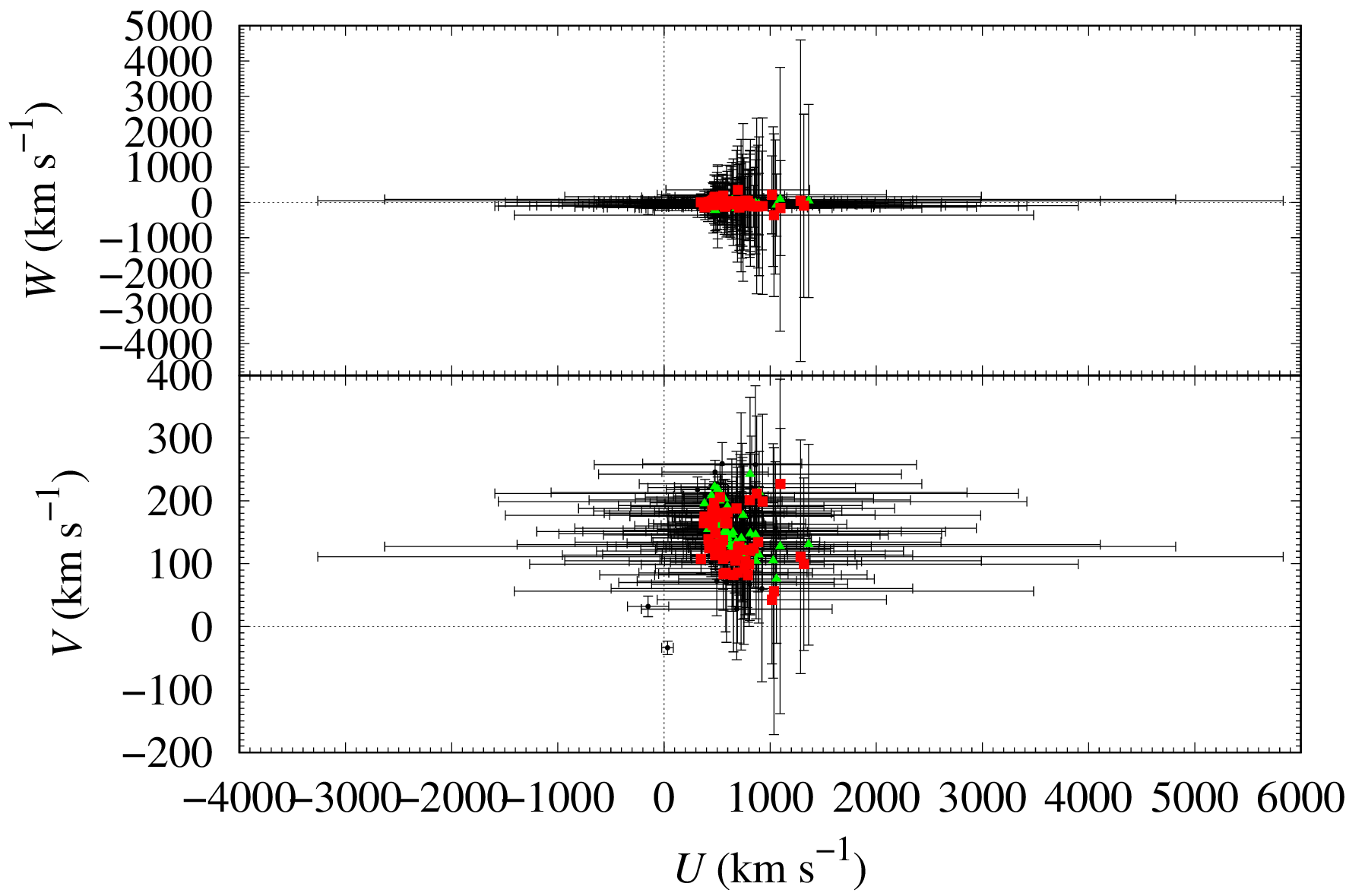}
         \includegraphics[width=\linewidth]{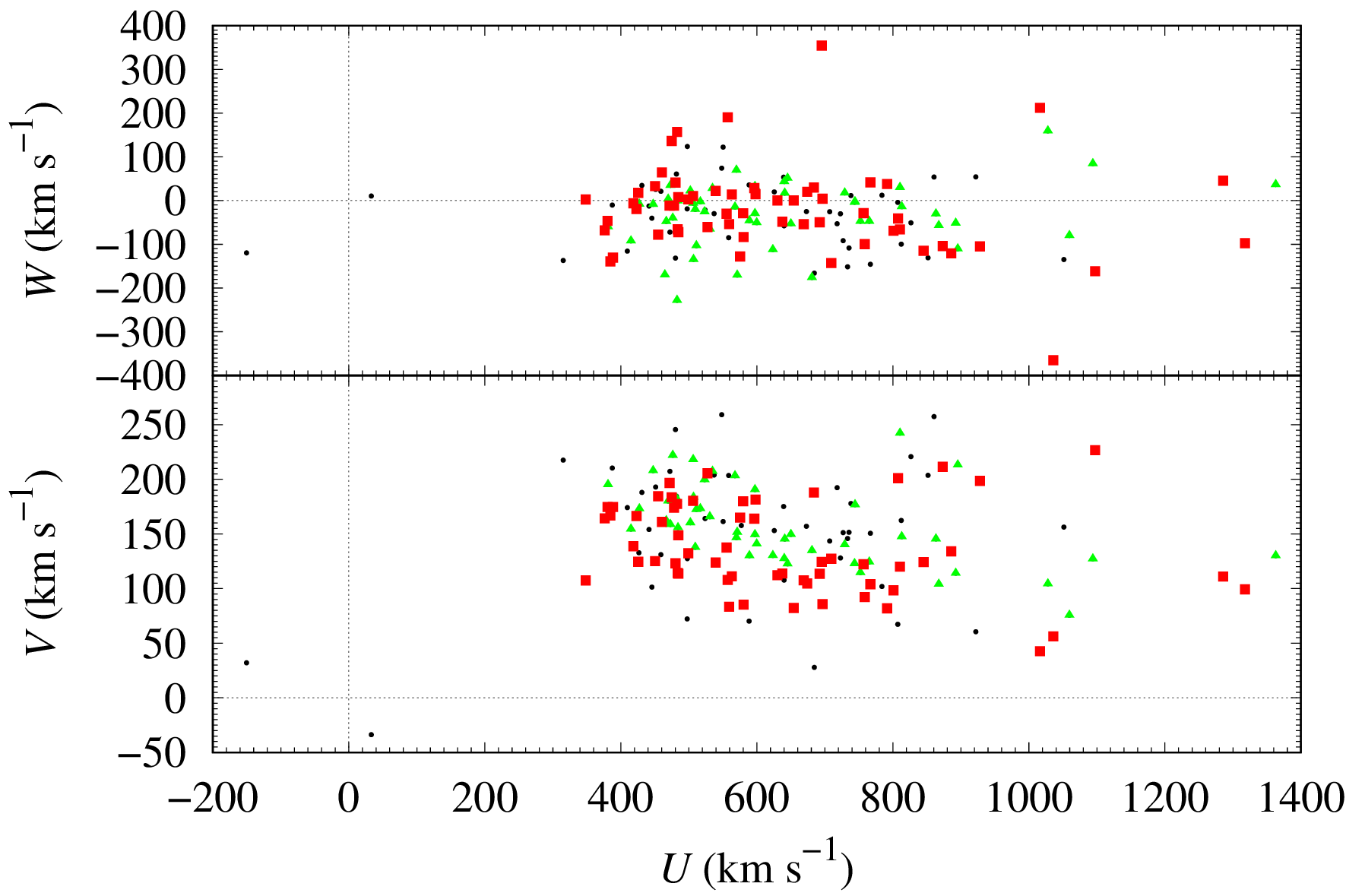}
         \caption{Galactocentric velocity components with (top two panels) and without error bars (bottom two panels). Triangles in green 
                  and squares in red as in Figs~\ref{pm}--\ref{full}.
                 }
         \label{galactovs}
      \end{figure}

     Fig.~\ref{galactovs} makes the issue of the uncertainties in the values of the distance explicit. In general, the Galactocentric 
     velocity components have very large error bars and this makes it virtually impossible to disentangle the two structures that may 
     contribute to the bimodal radial velocity distribution observed in Fig.~\ref{vrdist}, bottom panel. The prominent feature with
     central value 97$\pm$10~km~s$^{-1}$, may signal the presence of the long sought progenitor of the Monoceros Ring, but the data might 
     also be compatible with the existence of an unrelated, previously unknown, kinematically coherent structure, such as a distant spiral 
     arm or a spur that connects two well-separated arms. In Fig.~\ref{full}, top panel, the distribution above the disc looks quite 
     stream-like, as opposed to cluster-like. The location in the sky of structure B in fig.~5 of \citet{2004MNRAS.348...12M} matches 
     somehow that of the feature discussed here, but the distance range appears to be different. The same can be said about the Argo 
     structure discussed by \citet{2006ApJ...640L.147R}.
%
%
%
%
  
  \section{Conclusions}
     In this Letter, we have investigated the plausibility of the predictions made by \citet{2018MNRAS.474.4584G} regarding the possible 
     location of the putative progenitor of the Monoceros Ring using data from {\it Gaia} DR2 and focusing on the distribution of radial
     velocities. A statistically robust feature in the radial velocity distribution of stars in Vela has been identified. Our conclusions 
     are: 
     \begin{enumerate}[(i)]
        \item Based on the distributions of radial velocities provided by {\it Gaia} DR2, the distant stellar populations located in the
              regions proposed by \citet{2018MNRAS.474.4584G} appear to be compatible with single populations (i.e. no kinematically 
              heterogenous samples) in all but one case, that of the region towards the constellation of Vela.
        \item We have identified a statistically significant bimodal Gaussian distribution towards Galactic coordinates ($l$, $b$) = 
              (271{\degr}, +2{\degr}), which is one of the present-day locations of the progenitor of the Monoceros Ring proposed by 
              \citet{2018MNRAS.474.4584G}. This feature has central values of the radial velocity of 60$\pm$7~km~s$^{-1}$ and 
              97$\pm$10~km~s$^{-1}$.
        \item The prominent feature found towards Vela may signal the presence of the long sought progenitor of the Monoceros Ring, but 
              the data might also be compatible with the existence of an unrelated, previously unknown, kinematically coherent structure. 
        \item Interstellar extinction may be a major obstacle to disentangle the true nature of the two remote populations that appear to 
              share the patch of sky around ($l$, $b$) = (271{\degr}, +2{\degr}). 
     \end{enumerate}

  \section*{Acknowledgements}
     We thank the anonymous referee for a particularly insightful and constructive review, and A.~I. G\'omez de Castro for comments and for 
     providing access to computing facilities; RdlFM thanks L. Beitia-Antero for extensive discussions on {\it Gaia} DR2 data. This work was 
     partially supported by the Spanish MINECO under grant ESP2015-68908-R. In preparation of this Letter, we made use of the NASA 
     Astrophysics Data System, the ASTRO-PH e-print server, and the SIMBAD and VizieR databases operated at CDS, Strasbourg, France. This 
     work has made use of data from the European Space Agency (ESA) mission {\it Gaia} (\url{https://www.cosmos.esa.int/gaia}), processed by 
     the {\it Gaia} Data Processing and Analysis Consortium (DPAC, \url{https://www.cosmos.esa.int/web/gaia/dpac/consortium}). Funding for 
     the DPAC has been provided by national institutions, in particular the institutions participating in the {\it Gaia} Multilateral 
     Agreement.

  \bsp
  \label{lastpage}
\end{document}